\documentclass[preprint,12pt]{art}

\usepackage{amsmath}
\usepackage{amssymb}
\usepackage{amsthm}

\begin{document}

\begin{frontmatter}

\title{Large-radius Holstein polaron and the problem of spontaneous symmetry breaking.}

\author{Lakhno V.D.}

\address{Institute of Mathematical Problems of Biology,
Russian Academy of Sciences, Pushchino, Moscow Region, 142290, Russia}
\ead{lak@impb.psn.ru}

\begin{abstract}
A translation-invariant solution is found for a large-radius Holstein polaron whose energy in the strong coupling limit is lower than that obtained by Holstein. The wave function corresponding to this solution is delocalized. A conclusion is made about the absence of a spontaneous symmetry breaking in the quantum system discussed.
\end{abstract}


\end{frontmatter}

\section{Introduction.}
\label{1}
The question of whether the problem solution should have the same symmetry as the Hamiltonian led Landau to recognize that a polaron should be treated as an electron moving in an ideal lattice where a spontaneously arising fluctuation traps the electron to form a stable self-trapped state \cite{lit1}. In a classical one-dimensional lattice (molecular chain), a Bloch electron will always lose its initial symmetry if  the lattice can be deformed by it.  For the case of a one-dimensional molecular crystal, this problem was first considered by Holstein in \cite{lit2} (modern results on Holstein polaron are presented in \cite{lit2-1-Emin}). If the atoms of the lattice are considered quantum-mechanically, this conclusion will not be valid any longer. In a quantum lattice the symmetry of the electron-phonon system is conserved if the interaction of an electron with the lattice determined by the interaction constant $g$ is not too strong. For the value of g exceeding some critical value, according to \cite{lit2}, the symmetry is broken and a self-trapped state is formed. The statement made in \cite{lit2} that in the strong coupling limit a self-trapped polaron state is bound to arise contradicts, however, to the fact that the total momentum of the electron-phonon system in an ideal translationally symmetrical chain should be conserved.
Since the total momentum of the system is commuted with the Hamiltonian, the operator of the momentum and the operator of the Hamiltonian should have the same set of eigen functions. However, the eigen functions of the total momentum operator are plane waves, i.e. delocalized states, while those of the Hamiltonian operator in the strong coupling limit are  localized wave functions of the self-trapped state. This contradiction was analyzed in \cite{lit3}-\cite{lit4-1-Gerlach} where it was shown that for all the values of the coupling constant, the states should be delocalized. In the case of Pekar-Froehlich polaron these solutions in the strong-coupling limit were obtained in papers \cite{lit5}-\cite{lit9}. According to \cite{lit5}-\cite{lit9}, in the strong coupling limit delocalized polaron states have a lower energy than localized ones which break the symmetry.

In this paper we apply the approach of \cite{lit5} to the problem of a large-radius strong-coupling Holstein polaron \cite{lit2}. We will show that in this case, as in the case of Pekar-Froehlich polaron, in the limit of large $g$, minimum is reached in the class of delocalized wave functions.

The paper is arranged as follows. In \S2 we outline the results obtained for a large-radius strong-coupling Holstein polaron. A conceptual analogy is drawn between a formation of a Holstein polaron with spontaneously broken symmetry and models of elementary particles.

In \S3 we construct a general translation-invariant theory for a Holstein Hamiltonian. The construction is based on the approach which was first developed by Tulub \cite{lit5} for Pekar-Froehlich polaron and implied exclusion of electron coordinates from the Hamiltonian at the very beginning of the theory development. Hence, the theory is automatically translation-invariant by virtue of its construction.

In \S4 this general theory is applied to the case of weak interaction between an electron and a phonon field. In this limit case the theory reproduces the well-known limit of weak coupling for the ground state energy of a Holstein polaron.

In \S5 the strong coupling limit is considered. In this limit, we obtain a lower value for the ground state energy of a translation-invariant polaron, than Holstein did \cite{lit2} for a polaron  with spontaneously broken symmetry. A conclusion is made that in the quantum case, solutions conserving the symmetry of the initial Hamiltonian should be realized.

In \S6 the problem of various  consequences of the results obtained is discussed.

\section{ Holstein polaron in the strong coupling limit. Broken translation invariance.}
\label{2}
According to \cite{lit10}, Holstein Hamiltonian in a one-dimensional molecular chain in a continuum limit has the form (Appendix A):
\numberwithin{equation}{section}
\begin{equation}
\label {eq.21}
\hat{H}=-\frac{\hbar^2\nabla ^2}{2m}+\sum_{k}{V_k(a_ke^{ikx}+a_k^+e^{-ikx})}+\sum_{k}{\hbar\omega ^o_ka^+_ka_k};
\end{equation}
$$V_k=\frac{g}{\sqrt N}, \ \ \ \omega ^o_k=\omega _0$$
where  $a^+_k,\;a_k$ are the phonon field operators, $m$ is an electron effective mass, $\omega_0$ is the optical phonons frequency, $N$ is the number of atoms in the chain.

For the following analysis it is convenient to present some results concerning Hamiltonian (2.1) in the strong coupling limit. In Holstein theory \cite{lit2}, as well as in Pekar one \cite{lit11} it is believed that the wave function of the ground state has the form:
\begin{eqnarray}
\label {eq.22}
\Psi=\psi(x)\Phi (q_1,\ldots,q_k,\ldots)
\end{eqnarray}
where $\Phi$ is the phonon wave function, $\psi(x)$ is the electron wave function independent of phonon variables $q_k$. The ground state energy is determined from the condition of the total energy minimum $E$:
\begin{eqnarray}
\label {eq.23}
E=T-\Pi, \ \ \ T=\frac{1}{2m}\int\left|\nabla\psi\right|^2dx,\ \ \ \Pi=\frac{g^2a}{\hbar\omega_0}\int\left|\psi\right|^4dx,
\end{eqnarray}
where $a$ is the lattice constant.
Let us introduce a scaled  transformation of the wave function $\psi (x)$ retaining its normalization:
\begin{eqnarray}
\label {eq.24}
\psi (x)=\left|\xi\right|^{1/2}\tilde{\psi}\left(\left|\xi\right|x\right)
\end{eqnarray}
As a result, with the use of (2.3) we rewrite (2.2) as:
\begin{eqnarray}
\label {eq.25}
E(\xi)=\left|\xi\right|^2T-\left|\xi\right|\Pi
\end{eqnarray}
Fig. 1 shows the dependence  $E(\xi)$.

\includegraphics{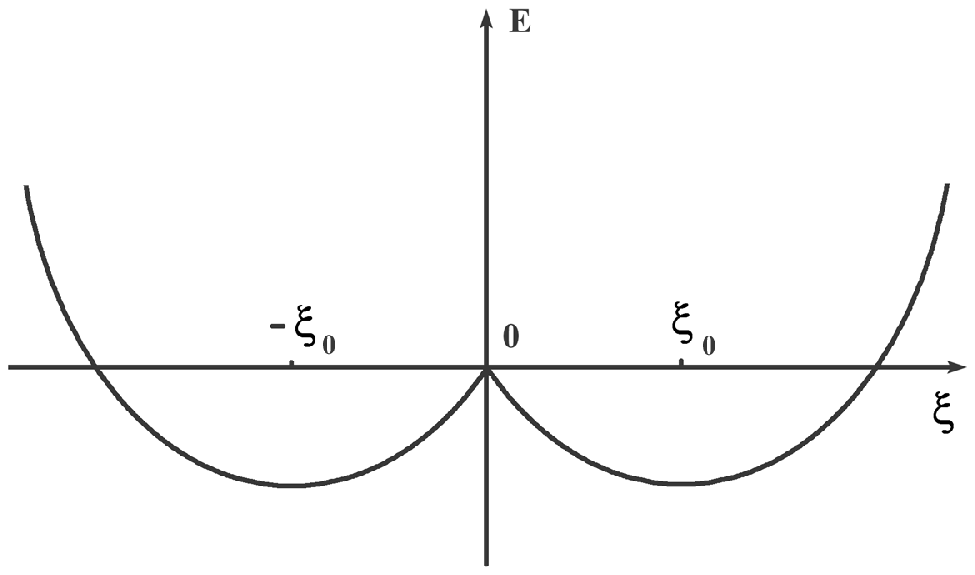}
\begin{center} Fig.1 \end {center}

Fig.1 illustrates why the global symmetry of the initially symmetrical delocalized state spontaneously breaks. The reason is that this state ($\xi=0$) corresponds to the local maximum of the functional $E$. The minimum of the functional at the points $\pm\xi_0$ corresponds to the energy and wave function $\psi(x)$:
\begin{eqnarray}
\label {eq.26}
E(\pm\xi_0)=-\frac{1}{6}\frac{\hbar^2}{mr^2},\ \ \psi(x)=\pm\left({\sqrt{2r}} \:\text{ch}\frac{x-x_0}{r}\right)^{-1},\ \
r=\frac{\hbar^3\omega}{mg^2a}
\end{eqnarray}
where $x_0$ corresponds to the position of the polaron well center in the energy minimum.

In the vicinity of this minimum one can carry out quantizing, restoring thereby the broken symmetry. Upon this restoration a Goldstone boson (zero phonon mode) automatically arises.

A similar approach is realized in models of elementary particles \cite{lit12}. For example, in the standard model not global (as in the case just considered), but local symmetry of gauge fields spontaneously breaks. In this case Goldstone bosons do not arise, while fields become massive .

In all the cases, in these models the symmetry spontaneously breaks and the restored one turns out to be less than the initial symmetry. Below, using Holstein model as an example, we will show that the approach discussed can lead to erroneous results.

\section{ Translation-invariant theory.}
\label{3}
In the previous section the results of the strong coupling theory with broken symmetry were given. Here we present the general translation-invariant approach reproducing the results of \cite{lit5} as applied to Holstein Hamiltonian.

In order to make the description translation-invariant let us exclude the electron coordinates from Hamiltonian (2.1)  using Heisenberg transformation \cite{lit13}, \cite{lit14}:
\begin{eqnarray}
\label {eq.31}
\hat{S}_1=\text{exp}\left\{\frac{i}{\hbar}\left(P-\sum_k{\hbar ka^+_ka_k}\right)x\right\},
\end{eqnarray}
where $P$ is the total momentum of the system. As a result of action of $\hat{S}_1$ on the field operators we get:
\begin{eqnarray}
\label {eq.32}
\hat{S}^{-1}_1a_k\hat{S}_1=a_ke^{-ikx},\ \ \ \hat{S}^{-1}_1a_k^+\hat{S}_1=a_k^+e^{ikx}.
\end{eqnarray}
Accordingly, the transformed Hamiltonian $\tilde{H}=S^{-1}_1HS_1$ takes the form:
\begin{eqnarray}
\label {eq.33}
\tilde{H}=\frac{1}{2m}\left(P-\sum_k{\hbar ka^+_ka_k}\right)^2+\sum_k V_k\left(a_k+a^+_k\right)+\sum_k \hbar\omega ^0_ka^+_ka_k
\end{eqnarray}
Let us subject the transformed Hamiltonian $\tilde{H}$ to one more canonical transformation \cite{lit15}:
\begin{eqnarray}
\label {eq.34}
\hat{S}_2=\text{exp}\left\{\sum_k f_k(a_k-a^+_k)\right\}
\end{eqnarray}
which leads to a shift of the field operators:
\begin{eqnarray}
\label {eq.35}
\hat{S}^{-1}_2a_k\hat{S}_2=a_k+f_k,\ \ \ \hat{S}^{-1}_2a_k^+\hat{S}_2=a_k^++f_k,
\end{eqnarray}
for real values of $f_k$. The resultant Hamiltonian $\tilde{\tilde{H}}=\hat{S}^{-1}_2\tilde{H}\hat{S_2}$  has the form:
\begin{eqnarray}
\label {eq.36}
\tilde{\tilde{H}}=H_0+H_1
\end{eqnarray}
where
\begin{eqnarray}
\label {eq.37}
H_0=\frac{P^2}{2m}+2\sum_kV_kf_k+\sum_k\left(\hbar\omega^0_k-\frac{\hbar kP}{m}\right)f^2_k+\\ \nonumber
\frac{1}{2m}\left(\sum_k kf^2_k\right)^2+\mathcal{H}_0
\end{eqnarray}
\begin{eqnarray}
\label {eq.38}
\mathcal{H}_0=\sum_k\hbar\omega _ka^+_ka_k+\frac{1}{2m}\sum_{kk'}kk'f_kf_{k'}(a_ka_{k'}+a^+_ka^+_{k'}+a^+_ka_{k'}+a^+_{k'}a_k)
\end{eqnarray}
\begin{eqnarray}
\label {eq.39}
H_1=\sum_k\left(V_k+f_k\hbar\omega_k\right)(a_k+a^+_k)+\frac{1}{2m}\sum_{kk'}kk'a^+_ka^+_{k'}a_ka_{k'}+\\\nonumber
\sum_{kk'}\frac{kk'}{m}f_{k'}(a^+_ka_ka_{k'}+a^+_ka^+_{k'}a_k),
\end{eqnarray}
\begin{eqnarray}
\label {eq.310}
\hbar\omega_k=\hbar\omega^0_k-\frac{\hbar k}{m}P+\frac{\hbar ^2k^2}{2m}+\frac{\hbar k}{m}\sum_k\hbar k'f^2_{k'}
\end{eqnarray}
In what follows we believe that $\hbar=1$, $m=1$.

According to \cite{lit5}, Hamiltonian (3.7) determines the ground state of the system, while Hamiltonian (3.9)
is eliminated by the proper choice of the wave function of the ground state.

Operator $\mathcal{H}_0$ is quadratic and can be reduced to a diagonal form. For this purpose we put:
\begin{eqnarray}
\label {eq.311}
q_k=\frac{1}{\sqrt{2\omega _k}}(a_k+a^+_k),\ \ \ \hat{p}_k=-i\sqrt{\frac{\omega _k}{2}}(a_k-a^+_k),\ \ \ z_k=kf_k\sqrt{2\omega _k}.
\end{eqnarray}
with the use of (3.11) expression (3.8) is written as:
\begin{eqnarray}
\label {eq.312}
\mathcal{H}_0=\frac{1}{2}\sum_k(p^+_kp_k+\omega ^2_kq^+_kq_k)+\frac{1}{2}\left(\sum_k z_kq_k\right)^2-\frac{1}{2}\sum_k\omega_k
\end{eqnarray}
Motion equations for operators $q_k$ and $p_k$:
\begin{eqnarray}
\label {eq.313}
\dot{q}_k=p_k\\\nonumber
\dot{p}_k=-\omega^2_kq_k-z_k\sum_{k'}z_{k'}q_{k'}
\end{eqnarray}
yield the following motion equation for $q_k$:
\begin{eqnarray}
\label {eq.314}
\ddot{q}_k+\omega^2_kq_k=-z_k\sum_{k'}z_{k'}q_{k'}
\end{eqnarray}
Let us seek a solution to system (3.14) in the form:
\begin{eqnarray}
\label {eq.315}
q_k(t)=\sum_{k'}\Omega_{kk'}\xi_{k'}(t),\ \ \  \xi_k(t)=\xi^0_ke^{i\nu_kt}
\end{eqnarray}
As a result the matrix $\Omega_{kk'}$ will be presented as:
\begin{eqnarray}
\label {eq.316}
(\nu ^2_k-\omega^2_k)\Omega_{kk'}=z_k\sum_{k''}z_{k''}\Omega_{k''k'}
\end{eqnarray}
Let us consider determinant of the system which is derived from (3.16) by replacing eigen values $\nu ^2_k$ by the quantity $s$ which can be
different from $\nu ^2_k$.

Determinant of this system will be:
\begin{eqnarray}
\label {eq.317}
\text{det}\left|(s-\omega^2_k)\delta_{kk'}-z_kz_{k'}\right|=\prod_k(s-\nu ^2_k)
\end{eqnarray}
On the other hand, according to \cite{lit16}:
\begin{eqnarray}
\label {eq.318}
\text{det}\left|(s-\omega^2_k)\delta_{kk'}-z_kz_{k'}\right|=\prod_k(s-\omega^2_k)\left(1-\sum_{k'}\frac{z^2_{k'}}{s-\omega^2_{k'}}\right)
\end{eqnarray}
It is convenient to introduce the quantity $\Delta(s)$:
\begin{eqnarray}
\label {eq.319}
\Delta(s)=\prod_k(s-\nu^2_k)/\prod_k(s-\omega^2_k)
\end{eqnarray}
From (3.17), (3.18) it follows that the quantity $\Delta(s)$ is equal to:
\begin{eqnarray}
\label {eq.320}
\Delta(s)=\left(1-\sum_k\frac{z^2_k}{s-\omega^2_k}\right)
\end{eqnarray}
The change in the system energy $\Delta E$ caused by the electron-field interaction is:
\begin{eqnarray}
\label {eq.321}
\Delta E=\frac{1}{2}\sum_k(\nu_k-\omega _k)
\end{eqnarray}
To express $\Delta E$ in terms of $\Delta (s)$ let us use Wentzel approach \cite{lit16}. Following \cite{lit16} we write down the identical equation:
\begin{eqnarray}
\label {eq.322}
\sum_k\left\{f(\nu^2_k)-f(\omega ^2_k)\right\}=\frac{1}{2\pi i}\oint_cdsf(s)\sum_k\left(\frac{1}{s-\nu^2_k}-\frac{1}{s-\omega^2_k}\right)=\\\nonumber
=\frac{1}{2\pi i}\oint_cdsf(s)\frac{d}{ds}\text{ln}\Delta(s)=-\frac{1}{2\pi i}\oint_cdsf'(s)\text{ln} \Delta(s)
\end{eqnarray}
where integration is performed over the contour, shown in Fig.2

\includegraphics{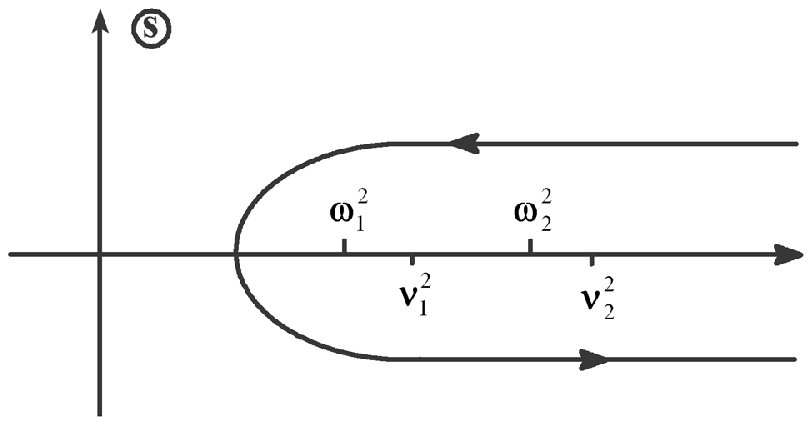}
\begin{center} Fig.2 \end {center}

Putting $f(s)=\sqrt{s}$, from (3.21), (3.22) we get:
\begin{eqnarray}
\label {eq.323}
\Delta E=\frac{1}{2}\sum_k(\nu_k-\omega _k)=-\frac{1}{8\pi i}\oint_c\frac{ds}{\sqrt{s}}\text{ln}\Delta(s)
\end{eqnarray}
Passing on in (3.20) from summing up to integration by using the relation:
\begin{eqnarray}
\label {eq.324}
\sum_k=\frac{1}{2\pi}\int dk
\end{eqnarray}
in the continuous case, expressing $z_k$ via (3.11) we write $\Delta(s)$ as:
\begin{eqnarray}
\label {eq.325}
\Delta(s)=D(s),\ \ \ D(s)=1-\frac{1}{\pi}\int^{\infty}_{-\infty}\frac{k^2f^2_k\omega_k}{s-\omega^2_k}dk
\end{eqnarray}
With the use of (3.23) and (3.25) in the particular case of $P=0$ we express the ground state energy of the operator $H_0$ as:
\begin{eqnarray}
\label {eq.326}
E=-\frac{1}{8\pi i}\oint_c\frac{ds}{\sqrt{s}}\text{ln}D(s)+2\sum_kV_kf_k+\sum_k f^2_k
\end{eqnarray}
The expression for the total energy $E$ determined by (3.26) is valid for the whole range of the coupling constant $g$ variation. In the next section we will consider the case of weak coupling.

\section{ The case of weak coupling}
\label{4}
The quantities $f_k$ in the total energy expression $E$ should be found from the minimum condition: $\delta E/\delta f_k$ which leads to the following integral equation for finding $f_k$:
\begin{eqnarray}
\label {eq.41}
f_k=-\frac{V_k}{(1+k^2/2\mu_k)},\ \ \ \mu^{-1}_{k}=\frac{\omega_k}{2\pi i}\oint_c\frac{ds}{\sqrt{s}}\frac{1}{(s-\omega^{2}_{k})D(s)}
\end{eqnarray}
In the case of weak coupling equations (4.1) can be solved using the perturbation theory. In the first approximation for $g\rightarrow0$ $D(s)=1$, the quantity $\mu^{-1}_{k}$ is equal to:
\begin{eqnarray}
\label {eq.42}
\mu^{-1}_{k}=\frac{\omega_k}{2\pi i}\oint_c\frac{ds}{\sqrt{s}}\frac{1}{(s-\omega^{2}_{k})}=1
\end{eqnarray}
Accordingly, from (4.1) $f_k$ will be expressed as:
\begin{eqnarray}
\label {eq.43}
f_k=-\frac{V_k}{1+k^2/2}
\end{eqnarray}
Hence, in the first approximation on $g$:
\begin{eqnarray}
\label {eq.44}
E=\Delta E+2\sum_kV_kf_k+\sum_kf^2_k
\end{eqnarray}
$$\Delta E=-\frac{1}{8\pi i}\oint_{c}\frac{ds}{\sqrt{s}}\text{ln}D(s)$$
$$\text{ln}D(s)\approx-\frac{1}{\pi}\int^{\infty}_{-\infty}\frac{k^2f^2_k\omega_k}{s-\omega^2_k}dk$$
Substituting (4.3) into (4.4) with regard to expression (1.1) for $V_k$, we transform (4.4) into a well-known (see for ex. \cite{lit10}, Appendix A) expression for the electron energy in the weak coupling limit:
\begin{eqnarray}
\label {eq.45}
E=-g^2\sqrt{ma^2/2\hbar^3\omega_0}
\end{eqnarray}
which we have written in dimensional units.

To calculate the other terms of the energy expansion in powers of $g$ we can use the approach developed in \cite{lit5}, \cite{lit17}.

\section{ The case of strong coupling.}
\label{5}
Now we pass on to a more complicated problem, i.e. that of a strong coupling approximation. To clear up the character of a solution in this region let us turn to analytical properties of the function $D(s)$ (3.25). For this purpose we present function $D(s)$  in the form:
\begin{eqnarray}
\label {eq.51}
D(s)=D(1)+\frac{s-1}{\pi}\int^{\infty}_{-\infty}\frac{k^2f^2_k\omega_kdk}{(\omega^2_k-1)(\omega^2_k-s)}
\end{eqnarray}
where $D(1)$ is the value of $D(s)$ for $s=1$:
\begin{eqnarray}
\label {eq.52}
D(1)=1+\frac{1}{\pi}\int^{\infty}_{-\infty}\frac{k^2f^2_k\omega_kdk}{\omega^2_k-1}\equiv 1+Q
\end{eqnarray}
Function $D(s)$, being a function of a complex variable $s$, has the following properties: 1) $D(s)$ has a crosscut along the real axis from $s=1$ to $\infty$ and has no other peculiarities; 2) $D^*(s)=D(s^*)$; 3) for $s\rightarrow\infty$, $sD(s)$ grows not slower than $s$. In view of these properties the function $\left[(s-1)D(s)\right]^{-1}$ can be presented in the form (see Appendix B):
\begin{eqnarray}
\label {eq.53}
\frac{1}{(s-1)D(s)}=\frac{1}{2\pi i}\oint_{c+\rho}\frac{ds'}{(s'-s)(s'-1)D(s')}
\end{eqnarray}
where $c+\rho$ is a contour shown in Fig. 3.

\includegraphics{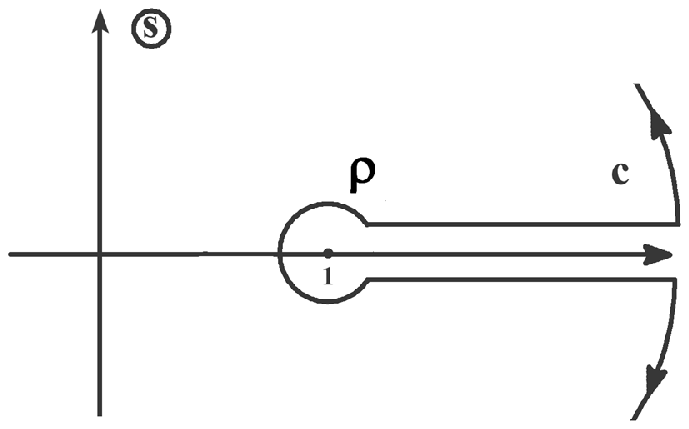}
\begin{center} Fig.3 \end {center}

From (5.3) follows that $D^{-1}(s)$ satisfies the integral equation:
\begin{eqnarray}
\label {eq.54}
\frac{1}{D(s)}=\frac{1}{1+Q}+\frac{s-1}{\pi}\int^{\infty}_{0}\frac{k^2f^2_k\omega_kdk}{(s-\omega^2_k)(\omega^2_k-1)\left|D(\omega^2_k)\right|^2}
\end{eqnarray}
whence with regard to (5.4) we get:
\begin{eqnarray}
\label {eq.55}
\Delta E=\frac{1}{4\pi}\int^{\infty}_{-\infty}\frac{k^2f^2_kdk}{2(1+Q)}+\\\nonumber
\frac{1}{4\pi^2}\int\int^{\infty}_{-\infty}\frac{k^2f^2_kp^2f^2_p\omega_p(\omega_k\omega_p+\omega_k(\omega_k+\omega_p)+1)}
{(\omega_k+\omega_p)^2(\omega^2_p-1)\left|D_+(\omega^2_p)\right|^2}dpdk
\end{eqnarray}
$$D_{\pm}(\omega^2_p)=1+\frac{1}{\pi}\int^{\infty}_{-\infty}\frac{k^2f^2_k\omega_kdk}{\omega^2_k-\omega^2_p\pm i\epsilon}$$
To calculate the polaron energy in the strong coupling limit let us choose the probe function $f_k$ in the form:
\begin{eqnarray}
\label {eq.56}
f_k=Nge^{-k^2/2a^2}
\end{eqnarray}
where $N$ and $a$ are variational parameters.
As a result, $\Delta E$ will be written as:
\begin{eqnarray}
\label {eq.57}
\Delta E=\frac{a^2}{32}(1+2q_T)
\end{eqnarray}
where $q_T$ is Tulub's integral \cite{lit5}:
\begin{eqnarray}
\label {eq.58}
q_T=\frac{2}{\sqrt{\pi}}\int^{\infty}_{0}\frac{e^{-y^2}(1-\Omega(y))dy}{v^2(y)+\pi y^2e^{-2y^2}/4}
\end{eqnarray}
$$\Omega (y)=2y^2\left\{(1+2y^2)ye^{y^2}\int^{\infty}_{y}e^{-t^2}dt-y^2\right\}$$
$$v(y)=1-ye^{-y^2}\int^{y}_{0}e^{t^2}dt-ye^{y^2}\int^{\infty}_{y}e^{-t^2}dt$$

for which an approximate value: $q_T\approx5,75$ was obtained in \cite{lit5} with a high degree of accuracy.

With the use of (3.26), (5.7), (5.6) the ground state energy takes the form:
\begin{eqnarray}
\label {eq.59}
E=\Delta E+2\sum_kV_kf_k+\sum_kf_k^2\approx\frac{12,5}{32}a^2-\sqrt{\frac{2}{\pi}}\left(1-\frac{N}{2\sqrt{2}}\right)g^2aN
\end{eqnarray}
Minimization of (5.9) by parameters $a$ and $N$ yields the following value of the polaron ground state energy:
\begin{eqnarray}
\label {eq.510}
E\approx-0,2037\frac{ma^2_0}{\hbar^2}\!\frac{g^4}{\hbar^2\omega^2_0}
\end{eqnarray}
which is presented in dimensional units. This result is fundamental since the energy value obtained is lower than that of Holstein polaron (2.6) (Appendix A).

For the probe function chosen in the form (5.6), the virial theorem holds: $2T=\Pi$, $W=3E$, where $W$ is the electron energy. Notice, that the mere fact that the virial theorem holds says nothing of whether the symmetry in the state under consideration is broken or not and for Holstein polaron, the virial theorem holds true both in the case of broken symmetry (\S2), and in the state with translation symmetry discussed here.

Equating the values of the weak coupling energy (4.5) and the strong coupling one (5.10) we can, as is customary in the polaron theory, get the value of the dimensionless coupling constant $g_c=g/\hbar\omega_0$ at which a transition from weak coupling to strong one occurs (Appendix A):
\begin{eqnarray}
\label {eq.511}
g_c\approx1,86\sqrt[4]{\hbar/ma^2_0\omega_0}
\end{eqnarray}
It should be stressed, however, that no jump-like transition from weak and intermediate coupling to the strong one occurs. The polaron state remains delocalized for all the values of the coupling constant and $E(g)$ is an analytical function $g$ \cite{lit3}, \cite{lit4}. This conclusion automatically results from the analytical properties of the function $D(s)$ too.

\section{ Discussion of the results.}
\label{6}
Turning back to the initial Landau's hypothesis that the electron wave function looses its symmetry because the electron forms a self-trapped state, we may state that this hypothesis is erroneous. This is seen even from Hamiltonian (3.3) which after Heisenberg canonical transformation (3.1) does not contain electron coordinates at all. Hence, the general form of the solution to Schroedinger equation for Hamiltonian (1) is:
\begin{eqnarray}
\label {eq.61}
\Psi=\text{exp}\left\{\frac{i}{\hbar}\left(P-\sum_k\hbar ka^+_ka_k\right)x\right\}\Lambda\left\{a_k\right\}\left|0\right\rangle
\end{eqnarray}
where $\Lambda\left\{a_k\right\}$ is a functional of the field operators, $\left|0\right\rangle$ is a vacuum wave function representing plane waves. It is important that electron and phonon variables are not separated in (6.1) as distinct from the case of broken symmetry (§2).

According to Fig.1, classically, this solution is unstable: under any infinitely small change of the classical orbit at this point, the amplitudes of the electron trajectory deviation will grow and tend to their finite values. Quantum-mechanical consideration widens the space of admissible states and leads to the possibility of stable oscillations in the vicinity of the classically unstable point $\xi=0$.

Quantum-mechanical states determined by the solution (6.1) do not have their classical analog. In particular, from (6.1) it follows that for the solution found, the notion of a classical polaron well localized in space is lacking, since a plane wave cannot support a finite value of atoms' displacements from their equilibrium positions.

Inapplicability of adiabatic approximation (2.2) in translation-invariant systems can also be qualitatively illustrated by the following reasoning. The criterion of applicability of adiabatic approximation is the smallness of the relation $m/M$,  where $M$ is the mass of the lattice's atom. For $m/M\rightarrow0$, i.e. when the mass of the atom tends to infinity, it can be considered as a classical particle. Accordingly, the field of displacements can be regarded as classical. Then separation of the motions determined by (2.2) becomes physically obvious: a localized electron described by the wave function $\psi(x)$, executing finite movements fast oscillates in the potential well and heavy atoms perceive only its averaged motion, having no time to fit into its position in space at each instant of time. In other words, an electron is presented as a static charge distributed with the density $\left|\psi(x)\right|^2$.

The physical situation changes if an electron is delocalized. In this case it executes infinite movements and even an infinitely heavy atom will have time to displace by a finite value by the moment when the electron occurs in its vicinity. Hence adiabatic approximation (2.2) in this case turns out to be invalid. Accordingly, the treatment of the displacements field as classical is invalid too.

The fact that the delocalized solution has a lower energy than the localized one has numerous physical consequences similar to those discussed in \cite{lit5}-\cite{lit9} for the case of Pekar-Froelich polaron. First of all, by analogy with an electron in an ideal rigid lattice in which Bloch electrons are superconducting ones, in a deformable lattice at zero temperature supercondacting particles are delocalized polarons described by wave function (6.1).  Any attempts to separate the electron-phonon system into a polaron and a phonon field in which a polaron is assumed to move in order to calculate the polaron mobility  \cite{lit18}, in view of (6.1), turn out to be unrealizable for a translation-invariant polaron. When a lattice has defects, translation-invariant polarons will not be trapped by them if the gain in energy provided by polaron's localization on a defect does not exceed that offered by the formation of a translation-invariant polaron. This is the qualitative difference between the translation-invariant polarons and Holstein ones with spontaneously broken symmetry since the latter will localize themselves on a defect at arbitrarily small value of the trap potential.

Since the global symmetry is preserved for translation-invariant polarons, Goldstone modes will be absent in their spectra, while for a Holstein polaron, the zero mode will always be present  in the phonon spectrum. Notice also, that in the case of translation-invariant polarons, the phonon spectrum will lack local modes arising in the course of formation of a Holstein polaron \cite{lit19}, \cite{lit20}.

In this case, for translation-invariant polarons, the energy values of renormalized frequencies of delocalized phonon modes $\nu _k$ are always higher than $\omega_k$, while for a strong coupling polaron with spontaneously broken symmetry, they are less than the value of $\omega_k$.

The foregoing proves that in order to get delocalized states preserving translation invariance, generally speaking, there is no need to break spontaneously the symmetry in the initial non-quantized Hamiltonian of the system, i.e. there is no need for a procedure suggested by Higgs so that to introduce the mass of elementary particles \cite{lit21}. It is also shown that investigation of extrema points of the relevant classical Hamiltonian cannot provide information of where (in the vicinity of what extremum) the quantum problem is to be solved. For the Holstein Hamiltonian considered, such an extremum is maximum of the classical Hamiltonian. The situation with Pekar-Froehlich polaron is similar \cite{lit6}-\cite{lit9}. The above statements, though referring to a nonrelativistic model, are, obviously, general since the polaron model is the simplest reach in content example of the quantum field theory.

In conclusion the author wishes to express his appreciation to A.V. Tulub for numerous discussions of the problems considered here.
The work was done with the support from the RFBR, Project N 13-07-00256.

\appendix
\section{}
\label{A}
Presently, Holstein Hamiltonian \cite{lit2} for an electron in a
homogeneous molecular chain is written as:
\begin{eqnarray}
\label{A.1}
    H=-\nu\sum_{n}\left(c_{n}^{+}c_{n+1}+c_{n+1}^{+}c_{n}\right)+g\sum_{n}c_{n}^{+}c_{n}\left(b_{n}^{+}+b_{n}\right)+\\\nonumber
    \sum_{n}\hbar\omega_0 \left(b_{n}^{+}b_{n}+\frac{1}{2}\right)\,,
\end{eqnarray}
 where $\nu$ is a matrix element of the
electron transition between neighboring sites, $c_{n}^{+}$,
$c_{n}$ are the operators of the birth and annihilation of an
electron on the n-th site, $b_{n}^{+}$, $b_{n}$ are the operators
of the birth and annihilation of oscillations quanta on the n-th
site.

In the case of weak coupling ($g/\hbar\omega_0\ll1$) solutions of
\eqref{A.1} are Bloch waves and the ground state energy in the second
order of perturbation theory has the form \cite{lit-A-Klamt}:
\label {A.2}

\begin{equation}
\label {A.2}
    E_{k}=-2\nu - \frac{g^2/\hbar\omega_0}{\sqrt{1+4\nu/\hbar\omega_0}}\,,
\end{equation}
where $a_0$ is the lattice constant.

For $g/\hbar\omega_0\gg1$ Holstein considered two limit cases : a
small radius polaron $(\nu\ll\hbar\omega_0)$ and a large-radius
one $(\nu\gg\hbar\omega_0)$\footnote{For the first time the
problem of the small radius polaron for T=0 was solved by
Tyablikov \cite{lit-A-Tyablikov}.}.

Solutions in the case of a small-radius polaron are
translation-invariant states :
\begin{equation}
\label {A.3}
    |\psi_{k}\rangle=\frac{1}{\sqrt{N}}\sum_{n}e^{ikn}c_{n}^{+}e^{g/\omega_0(b_{n}^{+}-b_{n})}|0\rangle\,,
\end{equation}
which correspond to the energy spectrum :
\begin{equation}
\label {A.4}
    E_{k}=-2\nu e^{-g^2/\hbar^2\omega_0^2}\cos
    ka_0-g^2/\hbar\omega_0\,.
\end{equation}

To pass on to the limit of a large-radius polaron, let us use in
\eqref{A.1} $|n\rangle\langle0|$, $|0\rangle\langle n|$ instead of
$c_{n}^{+}$, $c_{n}$ and operators of the birth and annihilation
of phonons with momentum k : $a_{k}^{+}$, $a_{k}$ instead of
$b_{n}^{+}$, $b_{n}$. To this end we will use the relations :
\begin{equation}
\label {A.5}
    \begin{aligned}
    c_{n}^{+}&=|n\rangle\langle0|\,,\ \ c_{n}=|0\rangle\langle n|\,,\\
    b_{n}^{+}&=\frac{1}{\sqrt{N}}\sum_{k}a_{k}^{+}e^{-ikna_0}\,,\
    \ b_{n}=\frac{1}{\sqrt{N}}\sum_{k}a_{k}e^{ikna_0}\,.
\end{aligned}
\end{equation}
As a result \eqref{A.1} takes on the form :
\begin{equation}
\label {A.6}
\begin{split}
    H&=-\nu\sum_{n}\left(|n\rangle\langle n+1|+|n+1\rangle\langle n|\right) + \\
    &+ \frac{g}{\sqrt{N}}\sum_{n,k}\left(a_{k}e^{ikna_0}+a_{k}^{+}e^{-ikna_0}\right)|n\rangle\langle
    n|+ \sum_{k}\hbar\omega_0a_{k}^{+}a_{k}\,.
\end{split}
\end{equation}
We will seek the solution of Schr\"{o}dinger equation with
Hamiltonian (A.6) in the form :
\begin{equation}
\label {A.7}
    |\Psi\rangle=\sum_{n}\psi_{n}|n\rangle\,.
\end{equation}
As a result, we express the averaged
Hamiltonian $\bar{H}=\langle \Psi|H|\Psi\rangle$ as :
\begin{equation}
\label {A.8}
\begin{split}
    \bar{H}&=-\nu\sum_{n}\left(\psi_{n}^{*}\psi_{n+1}+\psi_{n}\psi_{n-1}^{*}\right)+ \\
    &+ \frac{g}{\sqrt{N}}\sum_{n,k}|\psi_{n}|^2 \left(a_{k}e^{ikna_0}+a_{k}^{+}e^{-ikna_0}\right)+\sum_{k}\hbar\omega_0a_{k}^{+}a_{k}\,.
\end{split}
\end{equation}

In the case of a large-radius polaron :
\begin{equation}
\label {A.9}
    \psi_{n\pm1}\approx\psi_{n}\pm\frac{\partial\psi_{n}}{\partial n
    a_0}a_0+\frac{1}{2}\frac{\partial^2\psi_{n}}{\partial(na_0)^2}a_0^2\,.
\end{equation}

Having introduced a continuous variable $x=na_0$ and passed on in
(A.8) from summation to integration, we get :
\begin{equation}
\label {A.10}
\begin{aligned}
    \bar{H}&=\int\Psi^{*}H\Psi dx\,,\\
    H&=-\frac{\hbar^2}{2m}\Delta_{x}+\frac{g}{\sqrt{N}}\sum_{k}\left(a_{k}e^{ikx}+a_{k}^{+}e^{-ikx}\right)+\sum_{k}\hbar\omega_0a_{k}^{+}a_{k}\,,
\end{aligned}
\end{equation}
where $m=\hbar^2/2\nu a_0^2$, i.e. Hamiltonian \eqref{eq.21}.

While in a discrete case the exact solution of Hamiltonian \eqref{A.1}
in the strong coupling limit is known and determined by formulae
\eqref{A.3}, \eqref{A.4}, in the continuum case the exact solution of
Hamiltonian \eqref{A.10} for $g/\hbar\omega_0\gg1$ is not known.

The energy value \eqref{eq.510} found in this paper for
$g/\hbar\omega_0\gg1$ is currently the lowest one (Fig.~4).
\begin{center}

\includegraphics{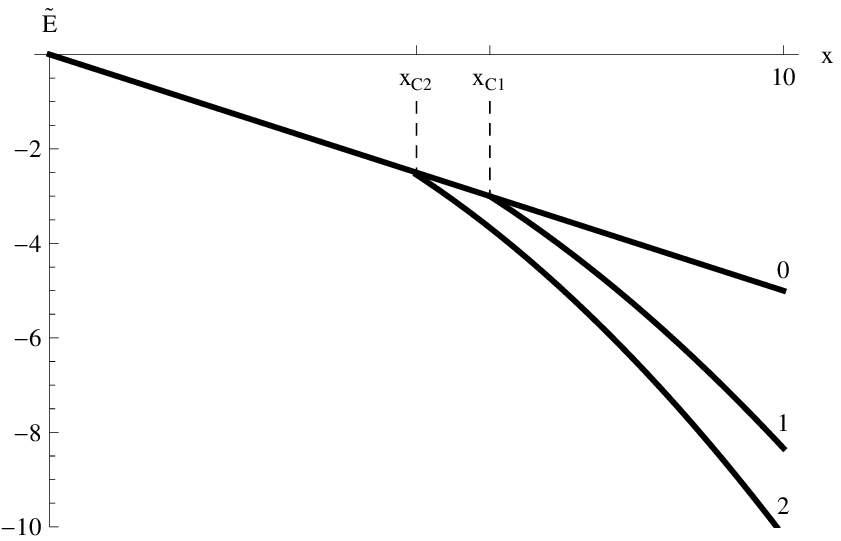}

Fig. 4. Dependence of energy $\tilde{E}=E/\hbar\omega_0$ on the
coupling constant $g_c$:\\

0) weak coupling in the continuum limit
$(\nu/\hbar\omega_0\gg1)$:\\ $\tilde{E}_{weak}=-0.5x$,
$x=\sqrt{\hbar\omega_0/\nu}g_c^2$,
$g_c=g/\hbar\omega_0$, (formula 4.5)\\

1) strong coupling:\\
$\tilde{E}_{strong}=-0.08333x^2$, obtained by Holstein (formula
2.6)

2) strong coupling:\\
$\tilde{E}_{strong}=-0.10185x^2$, obtained in this paper (formula
5.10).\\
The values $x_{C1}\approx6$ and $x_{C2}\approx5$ correspond to
$g_{C1}\approx2.45\sqrt[4]{\nu/\hbar\omega_0}$ and
$g_{C2}\approx2.2\sqrt[4]{\nu/\hbar\omega_0}$ (formula 5.11)

\end{center}

It is of interest to compare asymptotic expressions (2.6), (5.10)
with some real systems. For Holstein energy (2.6), such a
comparison was made in computational experiments with a classical
molecular chain as applied to DNA [26]. The results are in good
quantitative agreement. A comparison with energy (5.10) can be
made only for a quantum molecular chain with the use, for example,
of a quantum Monte Carlo method. Discussion of this problem,
however, beyond the scope of this paper.

\section{}
\label{B}

Let us show that (5.3), (5.4) follow from (5.1), (5.2).

To do this, we notice that analytical properties of $D(s)$ indicated in  \cite{lit5} straightforwardly follow from expression (3.25). Indeed, a pole corresponding to $D(s)$ can occur only on the real axis, since in view of positive definiteness $\omega_kk^2f^2_k$ in (3.25) equation:
\begin{eqnarray}
\label {eq.B1}
1+\frac{1}{\pi}\int^{\infty}_{-\infty}\frac{\omega_kk^2f^2_k(\omega_k-s_0+i\epsilon)}{(\omega^2_k-s_0)^2+\epsilon^2}dk=0
\end{eqnarray}
can be satisfied only for $\epsilon=0$. Besides, $D(s)$ is a monotonously increasing function of $s$ ($D'(s)>0$ for $s<1$) and for $s_0\rightarrow-\infty$ $D(s)$ turns to unity. Therefore $D(s)$ cannot have any zeros for $-\infty<s_0<1$.

On this basis, function $(s-1)D(s)$ can be presented in the form:
\begin{eqnarray}
\label {eq.B2}
\frac{1}{(s-1)D(s)}=\frac{1}{2\pi i}\int _{c+\rho}\frac{ds'}{(s'-s)(s'-1)D(s')}
\end{eqnarray}
where in Caushy integral in \eqref{eq.B2}  the contour of integration $c+\rho$ is shown in Fig. 3.

The integrand function in \eqref{eq.B2} has a pole for $s'=1$ and a crosscut from $s'=1$ to $s'\rightarrow\infty$. Performing integration in \eqref{eq.B2} along the upper and the bottom sides of the crosscut we will get integral equation (5.4).

\end{document}